# Incorporating effects of surface roughness and probing molecule size for estimation of soil specific surface area

Behzad Ghanbarian[1*], Allen G. Hunt[2,3], Marco Bittelli[4], Markus Tuller[5] and Emmanuel Arthur[6]

[1] Porous Media Research Lab, Department of Geology, Kansas State University, Manhattan KS 66506, United States of America

[2] Department of Physics, Wright State University, Dayton OH 45435, United States of America

[3] Department of Earth and Environmental Science, Wright State University, Dayton OH 45435, United States of America

[4] Department of Agricultural and Food Sciences, University of Bologna, Bologna, Italy

[5] Department of Environmental Science, The University of Arizona, Tucson AZ 85721, United States of America

[6] Department of Agroecology, Faculty of Technical Sciences, Technology Aarhus University, Tjele, Denmark

[*] Corresponding author's email address: ghanbarian@ksu.edu

## Abstract

The pore-solid interface and its characteristics play a key role in chemical interactions between minerals in the solid soil matrix and the liquid in pore space and, consequently, solute transport in soils. Specific surface area (SSA), typically measured to characterize the pore-solid interface, depends not only on the particles size distribution, but also particle shapes and surface roughness. In this note, we investigate the effects of surface roughness and probing molecule size on SSA estimation, employ concepts from fractals, and theoretically estimate specific surface area from particle size distribution and water retention curve (WRC). The former is used to characterize the particle sizes and the latter to approximately quantify the pore-solid interface roughness by determining the surface fractal dimension $D_s$. To evaluate our approach, we use five Washington and twenty one Arizona soils for which both particle size distributions and water retention curves were accurately measured over a wide range of particle sizes and matric potentials. Comparison with the experiments show that the proposed method estimates the SSA reasonably well with root mean square error RMSE = 16.8 and 30.1 $m^2/g$ for the Washington and Arizona datasets, respectively.

## 1. Introduction

### 1.1. Specific surface area measurement

In soils, chemical interactions between liquid in the pore space and minerals in the solid matrix, reactive transport, dissolution and deposition of minerals, and sorption are mainly controlled by the pore-solid interface and its characteristics. In numerous studies specific surface area (SSA) was measured to characterize the pore-solid interface. While its concept seems straightforward, in practice its accurate measurement and modeling have been challenging in different areas e.g., chemistry, chemical engineering, hydrology, soil science, and geotechnical engineering. Various methods, such as ethylene glycol monoethyl ether (EGME), Brunauer-Emmett-Teller (BET) isotherm, methylene blue (MB) titration, and MB-spot test have been proposed to experimentally measure specific surface area (Lowell and Shields, 2013). However, different methods can give widely different measurements of specific surface area (Sokołowska et al., 2001). For example, de Jong (1999) reported specific surface area measured by the EGME method 7 to 52 times greater than that measured by the $N_2$ adsorption method. Other ranges e.g., 1.4-35 and 2-9.2 were reported respectively by Yukselen and Kaya (2006) and Arthur et al. (2013). This might be due to the fact that $N_2$ adsorption only measures external surfaces (Carter et al., 1986), while EGME captures both internal and external surfaces (Yukselen and Kaya, 2006). In general, polar fluids are used to determine total SSA (internal and external) and non-polar fluids only capture external surfaces (Pennell, 2002). Carter et al. (1986) also stated that absolute measurement of specific surface area is difficult to determine due to the interaction of factors e.g., adsorbed cation, adsorbed molecule orientation, soil sample water content, etc.

In addition to $N_2$ (Pennell, 2002) and $CO_2$ (de Jonge and Mittelmeijer-Hazeleger, 1996), water vapor was also used to study sorption and to estimate SSA in soils (Arthur et al., 2016, 2014; Knadel et al., 2020; Leão and Tuller, 2014). For example, Arthur et al. (2016) evaluated nine different water vapor sorption models (including theoretical and empirical) using 207 soil samples with a wide range of texture. They found that for the adsorption isotherm the empirical Double Log Polynomial (Condon, 2006) model and for the desorption isotherm the Peleg model (Peleg, 1993) well characterized experimental measurements.

Arthur et al. (2013) measured adsorption and desorption isotherms for 21 soil samples from Arizona and determined the value of SSA from water sorption, EGME, and $N_2$-BET methods. They found that SSA calculated from water sorption was highly correlated to those determined from the EGME and $N_2$-BET methods with $R^2 = 0.94$ and 0.73, respectively. They also found the adsorption SSA to be smaller than the desorption SSA (hysteresis effect).

## 1.2. Specific surface area estimation

In the literature, different approaches such as artificial neural networks (Bayat et al., 2015; Knadel et al., 2020) and regression analysis (Chen et al., 2020; Ersahin et al., 2006; Hepper et al., 2006) were used to estimate specific surface area from other soil properties. For example, by analyzing 22 soil samples, Ersahin et al. (2006) proposed

$$SSA\left(\frac{m^2}{g}\right) = -6.56 + 3.96 Clay\% \tag{1}$$

with $R^2 = 0.53$. High correlation ($R^2 > 0.83$) between clay content and SSA was also reported by Resurreccion et al. (2011), Arthur et al. (2013), Leão and Tuller (2014), and Zinn et al. (2017).

In another study, using 24 samples, Hepper et al. (2006) empirically found

$$SSA\left(\frac{m^2}{g}\right) = 1.6925(Clay\% + Silt\%) - 10.552 \tag{2}$$

with $R^2 = 0.79$.

In addition to regression-based relationships, theoretical models were developed to estimate specific surface area from particle size distribution (Borkovec et al., 1993; Hunt et al., 2014; Hunt and Gee, 2002). For example, Hunt and Gee (2002) used a power-law probability distribution function to represent particle size distribution. They defined specific surface area as the surface area per volume of smooth spherical particles and developed a theoretic model to estimate specific surface area. However, Hunt and Gee (2002) did not compare their model estimations with experimental measurements. In another study, Koptsik et al. (2003) proposed a theoretical model to estimate total surface area from particle size distribution and an average surface form factor whose value for a spherical particle is 1. In their study, the surface form factor reflects both particle form and surface roughness, and its value for particles would be higher than that for smooth spheres. Since there were no data for explicit calculation of the average surface form factor, Koptsik et al. (2003) assessed its value by comparing estimated and measured surface areas.

## 1.3. Fractal characteristics of rough surfaces

It is well documented in the literature that the pore-solid interface in soils is rough and follows fractal properties (Anderson et al., 1996; Dathe et al., 2001; Gimenez et al., 1997; Pachepsky et al., 1996). While the surface roughness is characterized by the surface fractal dimension ($D_s$), the height asperity, or thickness, of the roughness is characterized by the root-mean-square of the roughness height (Cousins et al., 2018). The surface fractal dimension in three dimensions falls between 2 and 3 (Pfeifer et al., 1983). $D_s = 2$ indicates a smooth surface, while $D_s$ near 3 corresponds to an extremely rough surface. Hajnos et al. (2000) studied soil pore surface properties in managed grassland using the adsorption of water vapor and found that the surface fractal dimension ranged between 2.75 and 2.85. In another study, Bartoli et al. (1999) determined the surface fractal dimension from mercury intrusion porosimetry and capillary pressure curve in six silty soil samples and reported 2.61 $\leq D_s \leq$ 2.73 (see their Table 1). Sokołowska et al. (2001) measured specific surface area and calculated the surface fractal dimension from nitrogen and water vapor adsorption isotherms for 32 soil samples. They observed $2.10 \leq D_s \leq 2.59$ and $2.02 \leq D_s \leq 2.37$ from nitrogen and water vapor adsorptions, respectively. Sokołowska et al. (2001) did not find any correlation between these two surface fractal dimensions, although nitrogen and water vapor specific surface area values were relatively highly correlated with R = 0.84. Millán et al. (2013) also determined the surface fractal dimension from water vapor adsorption isotherms using a truncated fractal model. They reported $2.24 \leq D_s \leq 2.92$ in light, $2.21 \leq D_s \leq 2.94$ in medium, and $2.18 \leq D_s \leq 2.97$ in heavy alluvial soils. Modern techniques, such as high-resolution X-ray computed micro-tomography and image analysis can also be used to determine the value of surface fractal dimension in two and three dimensions (Dathe and Baveye, 2003; Katz and Thompson, 1985; Radlinski et al., 1999). However, capturing small pores and their roughness characteristics in soils remains challenging.

### 1.4. Objectives

Specific surface area depends on the method of measurement and adsorbate used to estimate it. Furthermore, the pore-solid interface in soils is typically rough with surface fractal dimension greater than 2. Accordingly, effects of the size of probing molecule (e.g., EGME, $H_2O$ or $N_2$) and surface roughness are non-trivial. Although there exist theoretical approaches in the literature to estimate the specific surface area, they are simplified models, or their parameters are generally unknown and/or have to be determined by matching estimated values with measured ones. Therefore, the main objectives of this study are to: (1) propose a method to estimate specific surface area from particle size distribution and surface fractal dimension, and (2) evaluate it experimentally using soil samples for which particle size distribution and water retention curve have been measured accurately over a wide range of particle sizes and matric potentials.

## 2. Theory

Results of David Avnir and his collaborators (Avnir et al., 1984, 1983; Pfeifer and Avnir, 1983) demonstrated that the rough surface of porous materials should follow fractal properties above some lower cutoff and below some upper cutoff scales. There might be some porous media that are (statistically) self-similar at any length scale. However, most natural materials showing fractal properties typically lose their fractal characteristics at sufficiently small or large length scales (Sahimi, 2003). Recently, Ghanbarian and Daigle (2015) highlighted that ignoring lower and upper cutoffs may result in inaccurate determination of fractal dimension.

Borkovec et al. (1993) argued that, in the case of rough particles, surface area increases with increasing particle radius and also with diminishing size of the probing molecule. Following Avnir et al. (1983), Borkovec et al. (1993) proposed that the surface area [L$^2$] of a particle with rough surface and radius $R$ [L] scales as:

$$SA = C_a \lambda^{2-D_s} R^{D_s} \tag{3}$$

where $D_s$ [-] is the surface fractal dimension, $C_a$ [-] is a constant coefficient, and $\lambda$ [L] is the size of probing (e.g., EGME or N$_2$) molecule. We should note that in the case of smooth spherical particles $D_s = 2$ and $C_a = 4\pi$. For a smooth surface ($D_s = 2$) the value of SA is independent of the size of adsorbing molecule, while for a rough surface the surface area increases with decreasing $\lambda$.

Since surface fractal dimension typically ranges between 2 and 3 (Avnir et al., 1985; Farin et al., 1985), surface area tends to infinity ($SA \to \infty$) as probing molecule size approaches zero $\lambda \to 0$, which is consistent with the Mandelbrot's definition of path lengths on geometrical fractals (Mandelbrot, 1982). The effect of probing molecule size on surface area for a rough pore-solid interface is schematically shown in Fig. 1 visualizing that as $\lambda$ decreases, $SA$ increases.

Combining Eq. (3) with the conceptual definition of specific surface area gives

$$SSA_v = \frac{\int_{R_{min}}^{R_{max}} c_a \lambda^{2-D_s} R^{D_s} f(R) dR}{\int_{R_{min}}^{R_{max}} s R^3 f(R) dR} = C_s \lambda^{2-D_s} \frac{\int_{R_{min}}^{R_{max}} R^{D_s} f(R) dR}{\int_{R_{min}}^{R_{max}} R^3 f(R) dR} \tag{4}$$

where $SSA_v$ [$L^{-1}$] is the volume-based specific surface area (surface area per unit of volume), $R_{min}$ and $R_{max}$ [L] are the minimum and maximum particle radii in the medium, $f(R)$ is the particle size distribution, $s$ [-] is a shape factor related to the volume of a grain, and $C_s = C_a/s$. In the case of smooth spherical particles, $s = \frac{4}{3}\pi$ and $C_a = 4\pi$, and, thus, $C_s = 3$ [-] (Borkovec et al., 1993). Eq. (4) is similar to Eq. (5.7) in Borkovec et al. (1993) who optimized the value of $D_s$ for one soil sample from surface area measurements on soil fractions of different sizes. They obtained a total specific surface area of 16.2 m$^2$/g, which compared favorably with the experimentally measured value of 18.8 m$^2$/g. We should point out that the Hunt and Gee (2002) model is a special case of Eq. (4) with $D_s = 2$ and $C_s = 1$. Eq. (4) indicates that the smaller the surface fractal dimension, the smoother the surface and consequently the smaller the specific surface area.

Eq. (4) returns $SSA_v$ in [$L^{-1}$]. However, one may use the following relationship to convert $SSA_v$ to the mass-based specific surface i.e., surface area per unit of mass ($SSA_m$) in [$L^2 M^{-1}$] (Knight and Nur, 1987; Sen et al., 1990):

$$SSA_m = SSA_v \frac{1}{\rho_s} \frac{\phi}{1-\phi} \qquad (5)$$

where $\rho_s$ is particle density [M L$^{-3}$] and $\phi$ is porosity [L$^3$ L$^{-3}$].

The estimation of specific surface area via Eq. (4) also requires the value of $D_s$, which can be determined from either image processing (Ogawa et al., 1999) or small angle x-ray scattering analysis (Borkovec et al., 1993). In this note, we propose to approximate the value of $D_s$ from water retention curve measurements. Ghanbarian et al. (2016) and Ghanbarian and Hunt (2017) also determined the value of surface fractal dimension from water retention curve and demonstrated that incorporating the effect of surface roughness improved unsaturated hydraulic conductivity estimations in soils.

The value of $C_s$ for rough grains in soils is a priori unknown. Accordingly, to estimate the SSA we set $C_s = 3$, a first-order approximation. Alternatively, its value can be determined by calibrating Eq. (4) using experimental measurements. By comparison with experiments, we demonstrate that this approximation ($C_s = 3$) leads to reasonable estimations of the SSA from measured $f(R)$ and estimated $D_s$ via Eq. (4). We should note that setting $C_s = 3$ does not necessarily mean that grains are smooth. In fact, Eq. (4) is more sensitive to the value of surface fractal dimension than the value of $C_s$. To demonstrate the sensitivity of Eq. (4) to the $D_s$ value, we compare the estimated SSAs using $D_s$ determined from the water retention curve and $D_s = 2$ (meaning that grains are smooth).

## 3. Materials and Methods

### 3.1. Washington soils

We used five Washington soil samples with three textures including sand, sandy loam and silt loam from Bittelli et al. (1999). Table 1 summarizes the salient properties of each soil sample. To measure particle size distribution in the range of 0.05 to 2000 μm, Bittelli et al. (1999) used a combination of wet sieving, pipette, and light-diffraction techniques. All the measured particle size distributions are shown in Fig. 2. Since the value of particle density is not available in this dataset, we used 2.65 g.cm$^{-3}$. The mass-based specific surface area was measured using the retention of ethylene glycol monoethyl ether (EGME) by Campbell and Shiozawa (1992) reported in the tabulated format by Tuller and Or (2005).

Water retention curves, shown in Fig. 3, were measured for the same five soil samples using a combination of the pressure plate and chilled-mirror dew point methods by Campbell and Shiozawa (1992). Each curve consists of between 31 and 39 paired measurements over a wide range of matric potentials i.e., between -3.3 × 10$^6$ and -3.1 × 10$^1$ cm H$_2$O, spanning five orders of magnitude.

### 3.2. Arizona soils

Twenty one Arizona soil samples were collected from the Arthur et al. (2013) study. This dataset includes a wide range of soil textures from coarse sand to clay (Table 1). A state-of-the-art Beckman Coulter LS 13320 laser diffraction particle size analyzer (Beckman Coulter Life Sciences, Indianapolis, IN, USA) that comprises of two laser systems – a standard system that covers the particle size range from 0.4 to 2000 μm and a polarization intensity differential scattering (PIDS) system that extends down to 0.04 μm while still providing a continuous size distribution up to 2000 μm – was applied to measure the particle size distribution between 0.04 and 2000 μm. To determine the PSD, a composite scattering pattern is measured by 126 detectors placed at angles up to approximately 35° from the optical axis. The light source is a 5-mW laser diode with a wavelength of 780 nm. For the PIDS system, a secondary tungsten-halogen light is projected through a set of filters that generates three wavelengths of 450, 600, and 900 nm. The Mie optical theory was applied to derive the particle size distributions from the measured diffraction patterns. Soil specific surface area was determined by means of the EGME method.

The wet-end water retention curves were measured with Tempe pressure cells (developed by the USDA in the city of Tempe, AZ). To capture the dry-end of the water retention curve, a WP4-T Dewpoint PotentiaMeter as well as an Aquasorb Vapor Sorption Analyzer (METER Group, Inc. USA) were used.

### 3.3. Surface fractal dimension determination

To determine the surface fractal dimension, the de Gennes (1985) model was fit to the measured water retention curves:

$$\theta = \phi \left( \frac{h}{h_a} \right)^{D_s - 3} \tag{6}$$

where $\theta$ [L$^3$ L$^{-3}$] is the water content, $\phi$ [L$^3$ L$^{-3}$] is the porosity, $h$ [L] is the matric potential, $h_a$ [L] is the air entry value, and $D_s$ is the surface fractal dimension. For a wide range of soils from the UNSODA database, Wang et al. (2005) showed that $D_s$ varied between 2 and 3. Eq. (6) is a special case of the more generalized model of Bird et al. (2000). Ghanbarian-Alavijeh and Millán (2009) showed that $D_s$ determined with Eq. (6) was highly correlated to water content retained at -15000 cm H$_2$O (known as permanent wilting point) with R$^2$ = 0.97 or to clay content with R$^2$ = 0.88. They stated that water content retained at permanent wilting point delineates the complex geometrical structure of the pore-solid interface.

To determine the values of $D_s$ and $h_a$ in Eq. (6), we first plotted water content versus matric potential on a log-log scale. Then, we detected the region between near full saturation and dry end of the water retention curve where the trend in the data was linear. Next, we fit Eq. (6) to that region (spanned several orders of magnitude in matric potential) and simultaneously optimized the value of $D_s$ and $h_a$. The fitted parameters of the de Gennes (1985) model i.e., $D_s$ and $h_a$ are reported in Table 2 for five Washington and 21 Arizona soil samples studied here.

### 3.4. Specific surface area estimation

For the sake of numerical integration and estimation of SSA, we used the Makima approach in MATLAB to interpolate between the measured data points on the particle size distributions. To estimate $SSA_v$ from the particle size distribution, we used the $D_s$ values calculated from the water retention curves and assumed $C_s = 3$, as a first-order approximation. Following Kellomäki et al. (1989), we set $\lambda = 0.75 \times 10^{-9}$ m ($= 0.75$ nm) for the EGME molecule size. We used the particle density and porosity values to convert the estimated $SSA_v$ to $SSA_m$ via Eq. (5).

### 3.5. Model evaluation criteria

To assess the reliability of the proposed SSA estimation method the log accuracy ratio (LAR) and root mean square error (RMSE) were determined as follows:

$$LAR = \log \left( \frac{SSA_{est}}{SSA_{meas}} \right) \tag{7}$$

$$RMSE = \sqrt{\frac{1}{N} \sum_{i=1}^{N} [SSA_{est} - SSA_{meas}]^2} \tag{8}$$

where $SSA_{est}$ and $SSA_{meas}$ are the estimated and measured specific surface areas, respectively, and $N$ is the number of samples. Note that LAR $= 0$ indicates a perfect match between estimated and measured SSA values.

## 4. Results and Discussion

Fig. 2 shows the measured particle size distributions for five Washington soils. As can be seen, particle radii span nearly 5 orders of magnitude covering a wide range from 0.01 to 1000 μm. Generally speaking, the measured particle size distributions are left-skewed indicating the presence of a heavy tail at smaller length scales. Some particle size distributions presented in Fig. 2 seem to show multimodal behavior. We numerically computed the integral in Eq. (4) using the actual measured particle size distribution. Accordingly, the multiscale characteristic of particle size distribution is not restrictive but implicitly considered for the estimation of the SSA. Similar results were obtained for 21 Arizona soil samples (not shown).

The fitted de Gennes (1985) model to the measured water retention curves are presented in Fig. 3 for five Washington soils. As can be seen, the model fits the data well with $R^2 \geq 0.95$ (Table 2). For the Washington dataset, we found the surface fractal dimension within a narrow range between 2.702 (Salkum) to 2.782 (L-soil). Interestingly, the optimized $D_s$ values obtained by fitting the de Gennes (1985) model are not much different from the mass fractal dimensions reported in Table 1 of Perfect (1999) who fitted a different form of water retention curve model to the same soil samples. We should point out that in the literature there are different fractal water retention curve models. For a recent comprehensive review see Ghanbarian and Millán (2017). Although proposed fractal models have different terminologies and interpretations for the fractal dimension, they might have the same mathematical forms. The ambiguity in the interpretation of water retention curve was raised by Crawford et al. (1995). They stated that, "A power-law exponent is a consequence of either a fractal pore volume; a fractal solid volume; a fractal pore wall; or a non-fractal, self-similar pore wall, and one cannot infer from the measurement which is the case." See Ghanbarian and Millán (2017) for further details and discussions.

For the Arizona dataset, surface fractal dimension ranged between 2.652 (Sample 1) and 2.868 (Sample 19), slightly wider than that for the Washington dataset. The $R^2$ values reported in Table 2 indicate that the de Gennes (1985) model fitted the measured water retention curves accurately with $R^2 \geq 0.95$. The average surface fractal dimension was 2.740 and 2.745 for the Washington and Arizona datasets, respectively. These values are greater than 2.4, the value determined from surface area measurements by Borkovec et al. (1993) for the Buchberg soil. However, our results are close to the results of Dathe et al. (2001) who reported $D_s \approx 2.6$ for the Gottingen soil. Dathe et al. (2001) investigated the effect of image resolution on the value of $D_s$ and found surface fractal dimension decreased from 2.9 to 2.6 as magnification increased.

In most soil samples analyzed here, the water content against the matric potential plotted on log-log scale showed one linear trend from near full saturation to oven dryness. This means one single surface fractal dimension could accurately scale the water retention curve. However, for several samples the slope of the $\theta$-$h$ curve changed at the dry end. In this region, the dominant mechanism is surface adsorption, and thus Eq. (6) may not accurately characterize the water retention curve. Thus, we excluded those measurements and determined the value of $D_s$ from the data points measured between near full saturation to almost oven dryness (on average ~300,000 cm $H_2O$). We should point out that if the water retention curve is not available, one may approximately estimate the value of $D_s$ from either the water content retained at -15000 cm H$_2$O or clay content (Ghanbarian-Alavijeh and Millán, 2009).

It has also been reported in the literature (Hunt et al., 2013; Millán and González-Posada, 2005) that surface roughness of soils may show multiscale characteristics. For example, when two surface fractal dimensions (i.e., $D_{s1}$ and $D_{s2}$) characterize the water retention curve with a crossover pore radius $r_x$, for estimating specific surface area Eq. (4) should change to

$$SSA_v = \frac{C_s}{\int_{R_{min}}^{R_{max}} R^3 f(R) dR} \left[ \lambda^{2-D_{s1}} \int_{R_{min}}^{R_x} R^{D_{s1}} f(R) dR + \lambda^{2-D_{s2}} \int_{R_x}^{R_{max}} R^{D_{s2}} f(R) dR \right]. \qquad (9)$$

In Eq. (9), $R_x$ represents the crossover particle radius at which the surface roughness scaling changes. Since both the particle size distribution and water retention curve are required to estimate the specific surface area, one may apply the Arya and Paris (1981) method to convert $r_x$, determined from the water retention measurements, to $R_x$.

We should point out that precise estimations of specific surface area requires accurate measurements of water retention curve. Bittelli and Flury (2009) and Solone et al. (2012) reported significant discrepancies between pressure plate and Dewpoint Potentiameter measurements at matric potentials smaller than -1000 cm H$_2$O. They found that water saturation determined with the pressure plate method at -15000 cm H$_2$O was several times greater than that measured with the dew point meter technique. This might be because of the loss of hydraulic continuity between the pressure plate and soil sample or lack of equilibrium at very low water saturations. Because of very low hydraulic conductivity, experiment time might be not adequately long enough to reach equilibrium conditions.

To demonstrate the importance of incorporating the effect of surface roughness, we report the specific surface area values estimated by Eq. (4) with $D_s = 2$ (smooth particles) in addition to those estimated using the optimized $D_s$ values from the measured water retention curves. We found that Eq. (4) with $D_s = 2$ substantially underestimated the specific surface area for all samples by several orders of magnitude. This is confirmed through the LAR values reported in Table 3. We also determined the value of root mean square error and found RMSE = 62.5 and 71.7 m$^2$/g for the Washington and Arizona datasets, respectively. The obtained results clearly show the importance of surface roughness for the specific surface area estimation. In addition, our results demonstrate that the proposed model, Eq. (4), is sensitive to the value of surface fractal dimension. This is in accord with the results of Ghanbarian-Alavijeh and Hunt (2012) who demonstrated that their unsaturated hydraulic conductivity model based on a combination of percolation and fractal theories was sensitive to the surface fractal dimension value. They showed that a small change in $D_s$ value may cause substantial underestimation in unsaturated hydraulic conductivity.

Although Eq. (4) with the optimized $D_s$ value underestimates the specific surface area by nearly 11 and 56% respectively for the Walla Walla and Royal samples in the Washington dataset, its calculated RMSE = 16.8 m$^2$/g is significantly less than 62.5 m$^2$/g. In the Arizona dataset, we found Eq. (4) with the optimized $D_s$ overestimated SSA by 140% in Sample 3 and underestimated SSA by 3% for Sample 17.

To address the effect of probing molecule size, we also estimated the specific surface area using $\lambda = 1.5$ nm and the optimized values of $D_s$. Results (not shown) indicated that the value of SSA$_m$ was underestimated for all samples, as expected. We found RMSE = 34.4 and 33.2 m$^2$/g for the Washington and Arizona datasets, respectively. The value of relative error ranged between -73 and -47% in the Washington dataset and between -66 and 47% in the Arizona dataset.

Fig. 4 shows the estimated mass-based specific surface area ($SSA_m$) using Eq. (4) against the measured value for five Washington and twenty one Arizona soil samples. Our results indicate that Eq. (4) with the optimized $D_s$ value estimates the specific surface area from the measured particle size distribution and water retention curve reasonably well with RMSE = 16.8 and 30.1 $m^2$/g (Fig. 4) for the Washington and Arizona datasets, respectively. These values are comparable to those reported by Arthur et al. (2018) and Knadel et al. (2020). For example, Arthur et al. (2018) compared the EGME SSA with the water-sorption-based one for soil samples dominated by kaolinite, illite and mixed clays, and smectites (see their Fig. 6). They reported RMSE values ranged between 11.6 and 61.2 $m^2$/g. In another study, Knadel et al. (2020) estimated SSA via partial least square, artificial neural network and support vector machine for 54 soil samples and found 27 < RMSE < 43 $m^2$/g (see their Fig. 5).

One advantage of theoretical models e.g., Eq. (4) over empirical models such as Eqs. (1) and (2) is in the former model parameters are physically meaningful, while in the latter constant coefficients are empirical. Schaap and Leij (1998) emphasized that regression-based models are database-dependent meaning that their performance may depend strongly on the data that were used for calibration and evaluation.

Although we set $C_s = 3$, as a first-order approximation, the obtained results are within an acceptable error margin (Fig. 4). Discrepancies between theoretical calculations and experimental measurements observed for other samples might be due to inaccurate estimation of the $C_s$ value. One should note that in media whose average particle geometry significantly deviates from spherical, our assumption $C_s = 3$ becomes invalid.

It is evident that soils with finer particles have larger specific surface area. It is well documented in the literature that specific surface area depends critically on the smallest particle radius $R_{min}$ (Borkovec et al., 1993; Koptsik et al., 2003). Wu et al. (1993), Bittelli et al. (1999) and many others have reported particle radius as small as several nanometers. Accordingly, inaccurate characterization of particle size distribution particularly as small length scales might result in SSA underestimations.

The difference between the measured and estimated SSA might also be due to inaccurate measurement of SSA using the EGME method. In the EGME method, like ethylene glycol, it is assumed that EGME covers all interlayer and external surfaces, which is difficult to prove (Carter et al., 1986). Furthermore, bubbles created on minerals may also lead to incomplete coverage of EGME in the interlayer spaces (Dowdy and Mortland, 1967). Carter et al. (1986) also pointed out that it is probable that EGME molecules are associated with exchangeable cations in thicknesses greater than required for a monomolecular layer, as has been reported for ethylene glycol (McNeal, 1964).

Supplementary investigation is required to further evaluate the proposed model using a broader range of differently textured soils. In this study, we estimated the value of surface fractal dimension from the water retention curve. Since $D_s$ can be determined from small angle X-ray scattering (SAXS) and μ-CT images, comparing water retention curve with SAXS and μ-CT imaging in the calculation of surface fractal dimension should also be investigated.

**5. Conclusion**

In this note, we proposed a method for the estimation of the specific surface area (SSA) from the particle size distribution and the water retention curve. The model assumed that the pore-solid interface is fractal, and its roughness can be characterized by the surface fractal dimension whose value ranges between 2 and 3. Since SSA can be measured with different methods and adsorbates, we incorporated the effect of probing molecule size. Comparison with five Washington and twenty one Arizona soils for which the particle size distributions and water retention curves were accurately measured for a wide range of grain sizes and matric potentials showed that the proposed approach estimated specific surface area reasonably well with RMSE = 16.8 $m^2$/g for the Washington and 30.1 $m^2$/g for the Arizona datasets. We found that both surface fractal dimension and probing molecule size have substantial impact on the SSA estimation.

**Acknowledgement**

BG acknowledges financial supports through faculty startup from Kansas State University.

**Figure Captions**

Fig. 1. Schematic of a rough particle surface covered by a monolayer of adsorbed probing molecules of decreasing size (top to bottom).

Fig. 2. The particle size distribution of five Washington soil samples. We used the Makima approach in MATLAB to interpolate between the data points. The textural properties of the soil samples are given in Table 1.

Fig. 3. The measured water content against the absolute value of matric potential for five Washington soil samples. The blue line represents the de Gennes (1985) model, Eq. (6), fitted to the measurements. The optimized parameters $D_s$ and $h_a$ as well as the correlation coefficient $R^2$ are reported in Table 2.

Fig. 4. The calculated mass-based specific surface area via our proposed method, Eq. (4), against the measured one for (a) five Washington and (b) twenty one Arizona soil samples studied here. The red dashed line indicates the 1:1 line.

Table 1. Textural properties for 26 soil samples from Washington (WA) and Arizona (AZ) studied here.

| State | Sample | Sand | Silt | Clay | Soil texture |
|-------|--------|------|------|------|--------------|
| WA | L-soil | 0.89 | 0.06 | 0.05 | Sand |
| | Royal | 0.54 | 0.32 | 0.15 | Sandy loam |
| | Salkum | 0.19 | 0.59 | 0.23 | Silt loam |
| | Walla Walla | 0.23 | 0.63 | 0.14 | Silt loam |
| | Palouse | 0.11 | 0.68 | 0.21 | Silt loam |
| AZ | 1 | 0.91 | 0.07 | 0.02 | Sand |
| | 2 | 0.96 | 0.03 | 0.01 | Sand |
| | 3 | 0.81 | 0.14 | 0.05 | Loamy sand |
| | 4 | 0.74 | 0.18 | 0.09 | Sandy loam |
| | 5 | 0.80 | 0.14 | 0.05 | Loamy sand |
| | 6 | 0.67 | 0.25 | 0.09 | Sandy loam |
| | 7 | 0.71 | 0.16 | 0.14 | Sandy loam |
| | 8 | 0.59 | 0.32 | 0.10 | Sandy loam |
| | 9 | 0.53 | 0.26 | 0.21 | Sandy clay loam |
| | 10 | 0.22 | 0.53 | 0.25 | Silt loam |
| | 11 | 0.37 | 0.46 | 0.17 | Loam |
| | 12 | 0.39 | 0.40 | 0.21 | Loam |
| | 13 | 0.66 | 0.15 | 0.20 | Sandy clay loam |
| | 14 | 0.58 | 0.15 | 0.27 | Sandy clay loam |
| | 15 | 0.51 | 0.22 | 0.27 | Sandy clay loam |
| | 16 | 0.04 | 0.73 | 0.23 | Silt loam |
| | 17 | 0.26 | 0.45 | 0.29 | Clay loam |
| | 18 | 0.09 | 0.40 | 0.51 | Silty clay |
| | 19 | 0.29 | 0.19 | 0.52 | Clay |
| | 20 | 0.49 | 0.21 | 0.30 | Sandy clay loam |
| | 21 | 0.71 | 0.20 | 0.09 | Sandy loam |

Table 2. The porosity and the optimized parameters of the de Gennes (1985) model, Eq. (6), for 26 soil samples studied here.

| State | Sample | Porosity $\phi$ (cm³/cm³) | Particle density (g/cm³) | de Gennes (1985) | | |
|-------|--------|------|------|-------|-------|-------|
| | | | | $D_s$ | $|h_a|$ (cm H$_2$O) | $R^2$ |
| WA | L-soil | 0.18 | 2.65 | 2.782 | 1.21 | 0.95 |
| | Royal | 0.35 | 2.65 | 2.708 | 20.6 | 0.98 |
| | Salkum | 0.48 | 2.65 | 2.702 | 135.9 | 0.99 |
| | Walla Walla | 0.39 | 2.65 | 2.744 | 30.5 | 0.99 |
| | Palouse | 0.44 | 2.65 | 2.763 | 33.2 | 0.99 |
| AZ | 1 | 0.36 | 2.63 | 2.652 | 2.3 | 0.98 |
| | 2 | 0.36 | 2.60 | 2.676 | 1.1 | 0.96 |
| | 3 | 0.38 | 2.59 | 2.714 | 7.1 | 0.97 |
| | 4 | 0.38 | 2.62 | 2.678 | 22.1 | 0.99 |
| | 5 | 0.37 | 2.63 | 2.698 | 16.2 | 0.97 |
| | 6 | 0.39 | 2.68 | 2.708 | 43.4 | 0.99 |
| | 7 | 0.39 | 2.59 | 2.758 | 13.2 | 0.99 |
| | 8 | 0.40 | 2.57 | 2.686 | 49.2 | 0.99 |
| | 9 | 0.43 | 2.61 | 2.737 | 63.2 | 0.99 |
| | 10 | 0.46 | 2.57 | 2.735 | 241.6 | 0.95 |
| | 11 | 0.42 | 2.51 | 2.696 | 91.7 | 0.99 |
| | 12 | 0.40 | 2.69 | 2.760 | 89.3 | 0.99 |
| | 13 | 0.40 | 3.08 | 2.778 | 19.5 | 0.98 |
| | 14 | 0.42 | 2.76 | 2.809 | 21.6 | 0.99 |
| | 15 | 0.43 | 2.58 | 2.811 | 24.5 | 0.99 |
| | 16 | 0.45 | 2.46 | 2.755 | 221.7 | 0.99 |
| | 17 | 0.43 | 2.41 | 2.778 | 95.3 | 0.99 |
| | 18 | 0.43 | 2.44 | 2.813 | 261.2 | 0.99 |
| | 19 | 0.45 | 2.61 | 2.868 | 37.9 | 0.99 |
| | 20 | 0.41 | 2.63 | 2.828 | 36.8 | 0.99 |
| | 21 | 0.38 | 2.62 | 2.700 | 19.0 | 0.99 |

$h_a$ is air entry value and $D_s$ is surface fractal dimension.

Table 3. The measured specific surface area (surface area per unit of mass; $SSA_m$) as well as the estimated value for five Washington soil samples studied here. LAR is the log accuracy ratio, Eq. (7).

| State | Soil | EGME $SSA_m$ ($m^2/g$) | $SSA_m$ ($m^2/g$) $D_s = 2$ | LAR | $SSA_m$ ($m^2/g$) optimized $D_s$ | LAR |
|---|---|---|---|---|---|---|
| Washington | L-soil | 25[*] | 0.0007 | -4.56 | 19.03 | -0.12 |
| | Royal | 45 | 0.0025 | -4.26 | 19.67 | -0.36 |
| | Salkum | 51 | 0.0109 | -3.67 | 41.19 | -0.09 |
| | Walla Walla | 70 | 0.0230 | -3.48 | 62.24 | -0.05 |
| | Palouse | 97 | 0.0099 | -3.99 | 73.03 | -0.12 |
| Arizona | 1 | 6.55 | 0.0023 | -3.46 | 9.86 | 0.18 |
| | 2 | 6.51 | 0.0036 | -3.26 | 15.16 | 0.37 |
| | 3 | 10.01 | 0.0026 | -3.59 | 24.11 | 0.38 |
| | 4 | 19.69 | 0.0029 | -3.84 | 15.73 | -0.10 |
| | 5 | 24.65 | 0.0041 | -3.78 | 21.48 | -0.06 |
| | 6 | 38.84 | 0.0033 | -4.07 | 24.15 | -0.21 |
| | 7 | 38.95 | 0.0031 | -4.10 | 45.75 | 0.07 |
| | 8 | 28.1 | 0.0036 | -3.89 | 19.82 | -0.15 |
| | 9 | 62.35 | 0.0034 | -4.26 | 40.25 | -0.19 |
| | 10 | 92.21 | 0.0105 | -3.94 | 57.71 | -0.20 |
| | 11 | 61.36 | 0.0107 | -3.76 | 33.96 | -0.26 |
| | 12 | 74.32 | 0.0032 | -4.36 | 47.74 | -0.19 |
| | 13 | 66.89 | 0.0026 | -4.40 | 50.84 | -0.12 |
| | 14 | 56.99 | 0.0031 | -4.27 | 90.02 | 0.20 |
| | 15 | 78.03 | 0.0036 | -4.33 | 105.10 | 0.13 |
| | 16 | 90.57 | 0.0092 | -3.99 | 70.76 | -0.11 |
| | 17 | 78.71 | 0.0041 | -4.28 | 76.36 | -0.01 |
| | 18 | 157.95 | 0.0090 | -4.25 | 134.04 | -0.07 |
| | 19 | 129.44 | 0.0041 | -4.50 | 238.80 | 0.27 |
| | 20 | 104.31 | 0.0034 | -4.49 | 120.79 | 0.06 |
| | 21 | 24.44 | 0.0027 | -3.95 | 20.30 | -0.08 |

[*] The mass-based specific surface area was measured using the retention of ethylene glycol monoethyl ether (EGME) by Campbell and Shiozawa (1992) reported in the tabulated format by Tuller and Or (2005).

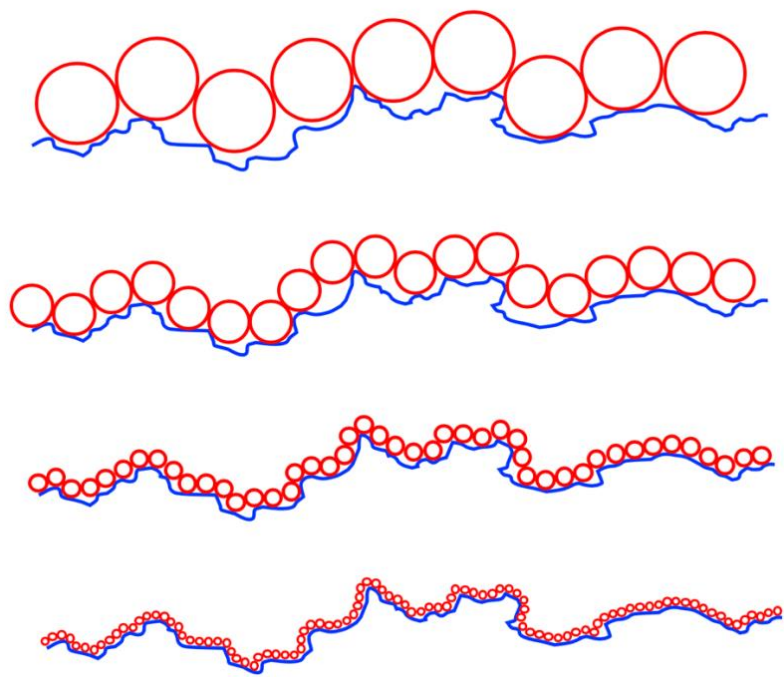

Fig. 1. Schematic of a rough particle surface covered by a monolayer of adsorbed probing molecules of decreasing size (top to bottom).

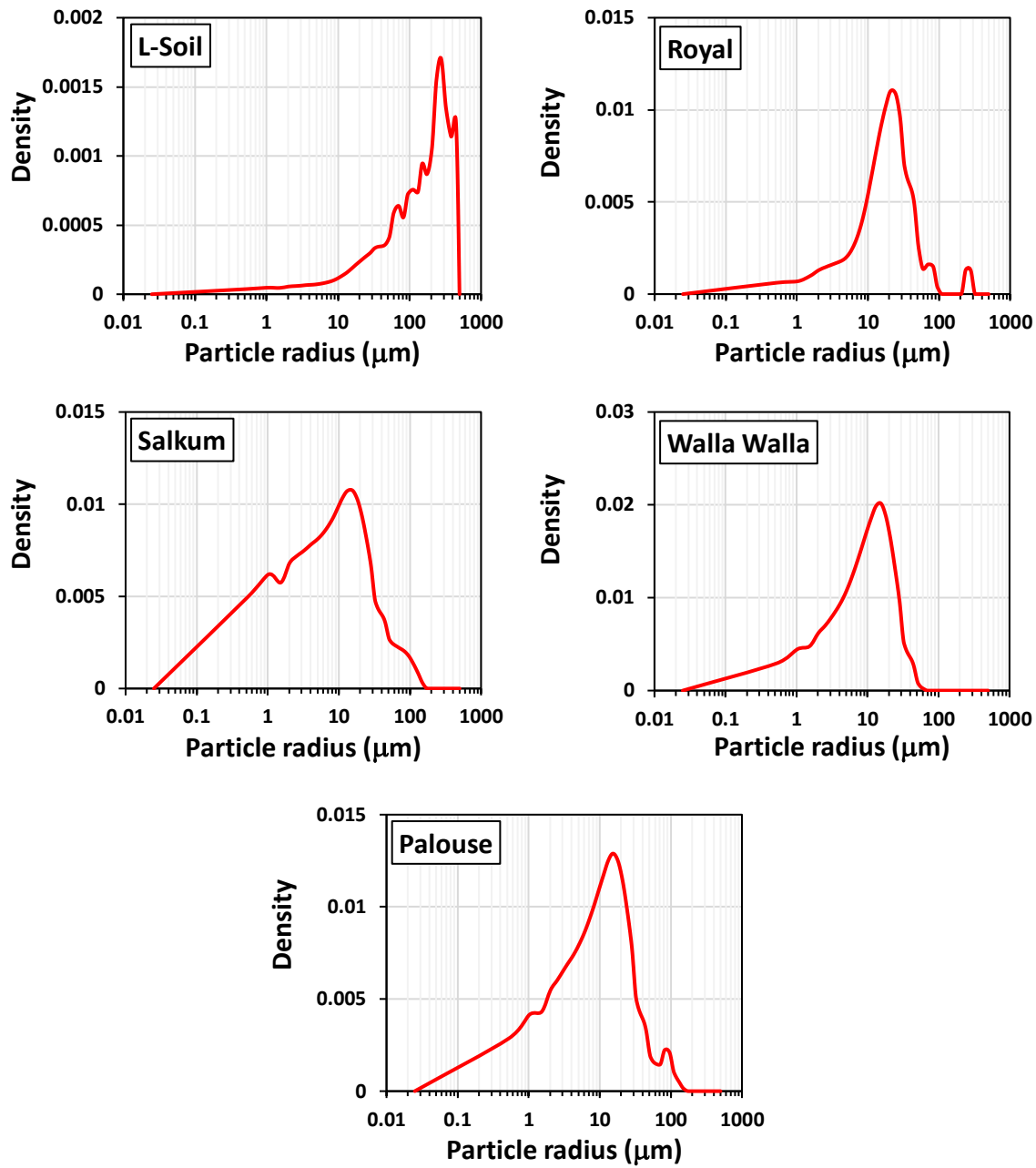

Fig. 2. The particle size distribution of five Washington soil samples. We used the Makima approach in MATLAB to interpolate between the data points. The textural properties of the soil samples are given in Table 1.

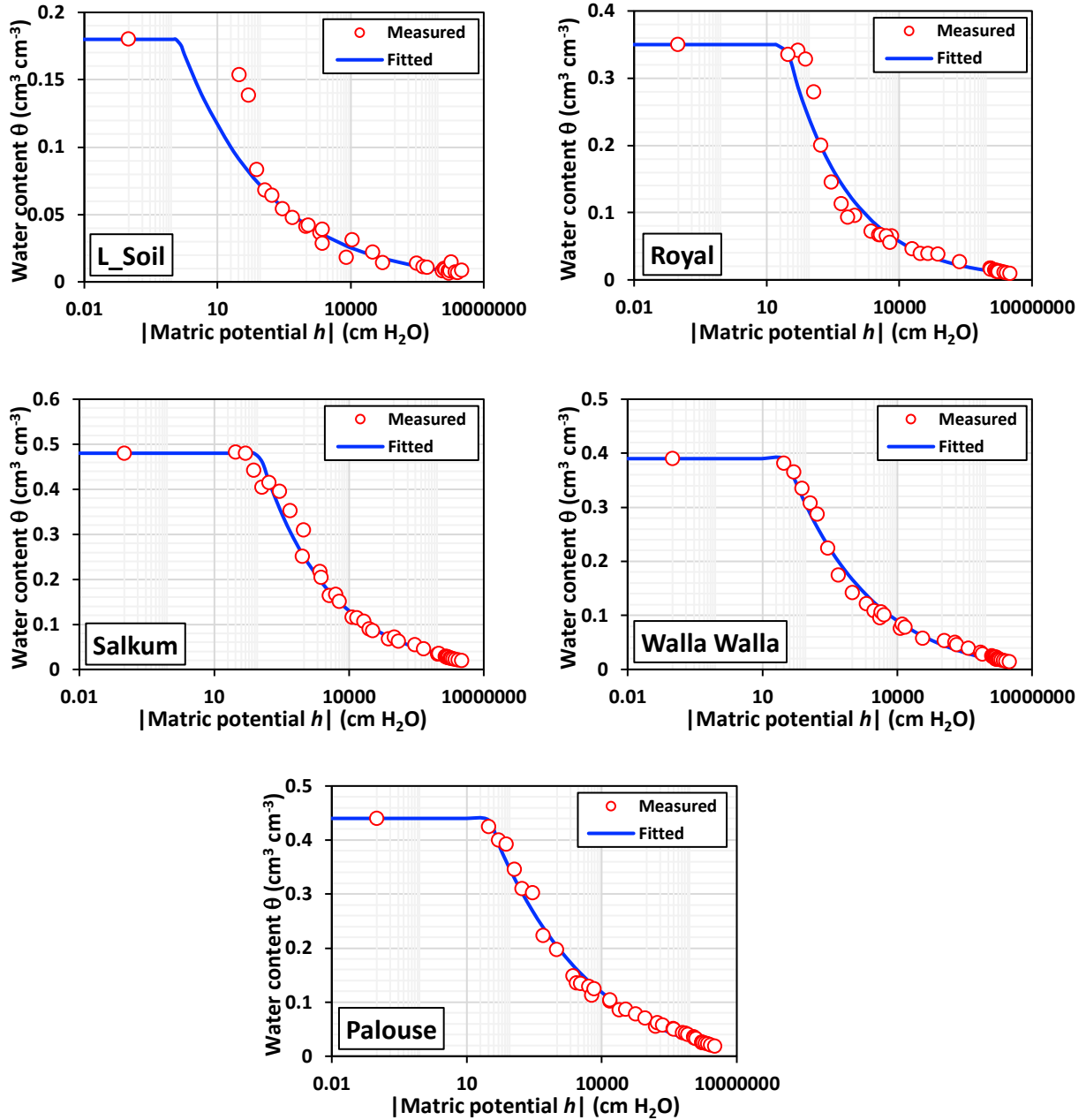

Fig. 3. The measured water content against the absolute value of matric potential for five Washington soil samples. The blue line represents the de Gennes (1985) model, Eq. (6), fitted to the measurements. The optimized parameters $D_s$ and $h_a$ as well as the correlation coefficient $R^2$ are reported in Table 2.

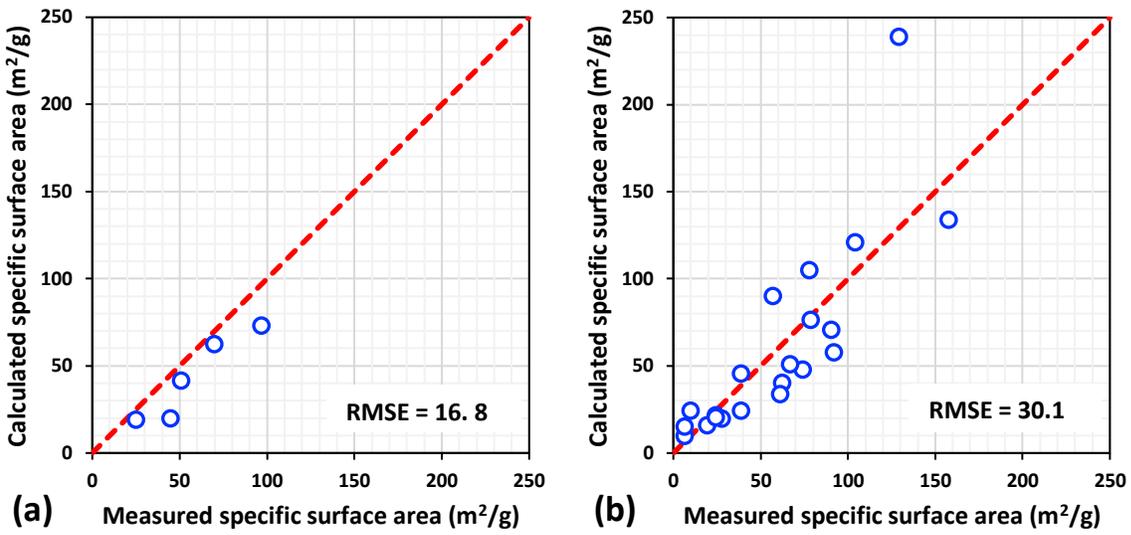

Fig. 4. The calculated mass-based specific surface area via our proposed method, Eq. (4), against the measured one for (a) five Washington and (b) twenty one Arizona soil samples studied here. The red dashed line indicates the 1:1 line.